\documentclass{stacs_proc}

\usepackage{latexsym} 
\usepackage{amssymb,amsmath}
\usepackage{epsfig}

\def\nat{{\mathbb N}}
\def\real{{\mathbb R}}

\newcommand{\sqrtsum}{{\tt{Sqrt-Sum}}}
\newcommand{\posSLP}{{\tt{PosSLP}}}

\begin{document}

\title[Equilibria, Fixed Points, and Complexity Classes]{Equilibria, Fixed Points, and Complexity Classes}

\author{Mihalis Yannakakis}{Mihalis Yannakakis}
\address{Department of Computer Science,  Columbia University, New 
York City, NY, USA}  
\email{mihalis@cs.columbia.edu}  

\thanks{Work supported by NSF Grant CCF-0728736.} 



\begin{abstract}
  \noindent Many models from a variety of areas involve the computation
of an equilibrium or fixed point of some kind. Examples include Nash equilibria
in games; market equilibria; computing optimal strategies and the values
of competitive games (stochastic and other games); 
stable configurations of neural networks; 
analysing basic stochastic models for evolution like branching processes 
and for language like stochastic context-free grammars;
and models that incorporate the basic primitives of probability and
recursion like recursive Markov chains. It is not known whether these problems
can be solved in polynomial time. There are certain common computational principles 
underlying different types of equilibria, which
are captured by the complexity classes PLS, PPAD, and FIXP.
Representative complete problems for these classes are respectively,
pure Nash equilibria in games where they are guaranteed to exist, 
(mixed) Nash equilibria in 2-player normal form games,
and (mixed) Nash equilibria in normal form games with 3 (or more) players.
This paper reviews the underlying computational principles and the corresponding classes.
\end{abstract}

\maketitle

\stacsheading{2008}{19-38}{Bordeaux}
\firstpageno{19}

\section{Introduction}

Many situations involve the computation of an equilibrium or a stable
configuration of some sort in a dynamic environment.
Sometimes it is the result of individual agents acting on their own noncompetitively
but selfishly (e.g., Nash and other economic equilibria), sometimes
it is agents acting  competitively against each other (and perhaps nature/chance),
sometimes the equilibrium is the limit of an iterative process that evolves
in some direction until it settles. Often the sought objects can be described
mathematically as the fixed points of an equation $x = F(x)$.

Many models and problems from a broad variety of areas are of this
nature.  Examples include: Nash equilibria in games; market
equilibria; computation of optimal strategies and the values of
competitive games (stochastic and other games); stable configurations
of neural networks; analysis of basic stochastic models for evolution
like branching processes, and for language like stochastic
context-free grammars; and models that incorporate the basic
primitives of probability and recursion like recursive Markov chains.
Most of these models and problems have been studied mathematically for
a long time, leading to the development of rich theories.  Yet, some
of their most basic algorithmic questions are still not resolved, in
particular it is not known whether they can be solved in polynomial
time.

Despite the broad diversity of these problems, there are certain common
computational principles that underlie many of these different types of problems,
which are captured by the complexity classes PLS, PPAD, and FIXP.
In this paper we will review these principles, the corresponding classes, 
and the types of problems they contain.

All the problems we will discuss are total search problems.
Formally, a {\em search problem} $\Pi$ has a set of instances, 
each instance $I$ has a set $Ans(I)$ of acceptable answers; 
the search problem is {\em total} if $Ans(I) \neq \emptyset$ 
for all instances $I$. 
As usual, for computational purposes, instances are represented by strings 
over a fixed alphabet $\Sigma$, and it is assumed
that, given a string over $\Sigma$ one can determine in polynomial
time if it represents an instance of a problem. The size $|I|$ of 
an instance is the length of its string representation.
Input numbers (such as the payoffs of games, input probabilities of stochastic models, etc.) 
are assumed to be rationals represented in binary by numerator and denominator.
The underlying solution space from which answers are drawn may be finite and discrete, 
as in combinatorial problems, 
or it may be infinite and continuous.
In the former (the finite) case, solutions are represented also as strings 
and the problem is: given an instance $I$, compute a solution in $Ans(I)$.
In the latter (infinite/continuous) case also, if there are rational-valued solutions
(as in Linear Programming for example), then the problem is to compute one of them.
In several problems however, the solutions are inherently irrational,
and we cannot compute them exactly (in the usual Turing machine-based
model of computation and complexity). In these cases we 
need to specify precisely which information about the
solutions is to be computed; this could be for example a
yes/no question, such as, does an event in a stochastic model occur almost surely (with
probability 1) or does the value of a game exceed a given threshold, or we
may want to compute an answer up to a desired precision.
In any case, the computational tasks of interest have to be defined precisely, 
because different tasks can have different complexity.

In this paper we will discuss a variety of equilibria and fixed point problems,
and the complexity classes which capture the essential aspects
of several types of such problems. We discuss three classes,
PLS, PPAD, and FIXP,
which capture different types of equilibria.
Some representative complete problems for these classes are:
for PLS pure Nash equilibria in
games where they are guaranteed to exist, for PPAD
(mixed) Nash equilibria in 2-player normal form games,
and for FIXP (mixed) Nash equilibria in normal form games with 3 (or more) players.

\section{Discrete, Pure Equilibria and the Class PLS}

Consider the following {\em neural network} model \cite{Ho}: 
We have an undirected graph $G=(V,E)$ with
a positive or negative weight $w(e)$ on each edge $e \in E$ 
(we can consider missing edges as having weight 0) and a threshold
$t(v)$ for each node $v \in V$. A configuration of the network is an
assignment of a state $s(v) =+1$ (`on') or $-1$ (`off') 
to each node $v \in V$. A node $v$ is {\em stable}
(or `happy') if $s(v)=1$ and $\sum_u w(v,u) s(u) + t(v) \geq 0$,
or $s(v)=-1$ and $\sum_u w(v,u) s(u) + t(v) \leq 0$,
i.e. the state of $v$ agrees with the sign of the
weighted sum of its neighbors plus the threshold.
A configuration is {\em stable} if all the nodes are stable.
A priori it is not obvious that such a configuration exists;
in fact for directed networks there may not exist any.
However, every undirected network 
has at least one (or more) stable configuration \cite{Ho}. 
In fact, a dynamic process where in each step one node that is unstable
(any one) switches its state, is guaranteed to eventually
converge in a finite number of steps to a stable configuration,
no matter which unstable node is switched in each step.
(It is important that updates be asynchronous, one node at a time;
simultaneous updates can lead to oscillations.)
To show the existence of a stable configuration and  convergence, 
Hopfield introduced a value function (or `potential' or `energy') on configurations,
$p(s)= \sum_{(v,u) \in E} w(v,u) s(v)s(u) + \sum_{v \in V} t(v)s(v)$. 
If $v$ is an unstable node in configuration $s$, then switching
its state results in a configuration $s'$ with 
strictly higher value $p(s') = p(s) + 2|\sum_u w(v,u) s(u) + t(v)| > p(s)$.
Since there is a finite number $(2^{|V|})$ of configurations, the process
has to converge to a stable configuration.
The {\em stable configuration problem} is the following: Given a neural network,
compute a stable configuration. This is a total search problem,
as there may be one or more stable configurations, and anyone of them
is an acceptable output. 

Although the stable configuration problem does not call a priori for
any optimization, the problem can be viewed equivalently as one of
{\em local} optimization: compute a configuration $s$ whose value
$p(s)$ cannot be increased by switching the state of any single node.
Local search is a common, general approach for tackling hard
optimization problems.  In a combinatorial optimization problem $\Pi$,
every instance $I$ has an associated finite set $S(I)$ of solutions,
every solution $s \in S(I)$ has a rational value or cost $p_I(s)$ that
is to be maximized or minimized.  In local search, a solution $s \in
S(I)$ has in addition an associated neighborhood $N_I(s) \subseteq
S(I)$; a solution is locally optimal if it does not have any
(strictly) better neighbor, i.e. one with higher value or lower cost.
A standard local search algorithm starts from an initial solution, and
keeps moving to a better neighbor as long as there is one, until it
reaches a local optimum.  The complexity class PLS (Polynomial Local
Search) was introduced in \cite{JPY} to capture the inherent
complexity of local optima for usual combinatorial problems, where
each step of the local search algorithm can be done in polynomial
time.  Even though each step takes polynomial time, the number of
steps can be potentially exponential, and in fact for many problems we
do not know how to compute even locally optimal solutions in
polynomial time.  Formally, a problem $\Pi$ is in PLS if solutions are
polynomially bounded in the input size, and there are polynomial-time
algorithms for the following tasks: (a) test whether a given string
$I$ is an instance of $\Pi$ and if so compute a (initial) solution in
$S(I)$, (b) given $I,s$, test whether $s \in S(I)$ and if so compute
its value $p_I(s)$, (c) given $I,s$, test whether $s$ is a local
optimum and if not, compute a better neighbor $s' \in N_I(s)$.
Notions of PLS reduction and completeness were introduced to relate
the problems.  A number of well-studied combinatorial optimization
problems (e.g. Graph Partitioning, TSP, Max Cut, Max Sat etc.)  with
common neighborhood structures (both simple and sophisticated) have
been shown to be PLS-complete by many researchers, and thus locally
optimal solutions can be computed efficiently for anyone of them iff
they can be computed for all PLS problems.  For a detailed survey and
bibliography see \cite{Ya2}.  In particular, the stable configuration
problem is PLS-complete (and is complete even if all thresholds are 0
and all weights are negative, i.e. all connections are repulsive)
\cite{SY}.

It is worth stressing several points: 1. The search problem asks to compute any
local optimum, not a specific one like the best, which is often NP-hard.
2. Given an instance $I$, we can always guess a solution $s$, and verify in
polynomial time that it is indeed a solution ($s \in S(I)$) and it is locally
optimal. Hence PLS is somewhere between P and TFNP (total search problems in NP).
Such problems cannot be NP-hard (under Cook reductions) unless NP=coNP.
3. We are interested in the inherent complexity of the search probleme itself
{\em by any algorithm whatsoever}, not necessarily the standard local search algorithm,
which often has exponential running time. For example, Linear Programming
can be viewed  as a local search problem (where local optima= global optima)
with Simplex as the local search algorithm; we know that Simplex under many
pivoting rules is exponential, yet the problem itself can be solved in polynomial
time by completely different methods (Ellipsoid, Karmakar).
In fact, many common local search problems are complete under a type of {\em tight} PLS-reduction
which allows us to conclude that the corresponding standard local search algorithm
is exponential. For example, in the neural network model, the dynamic
process where unstable nodes switch iteratively their state until the network
stabilizes takes for some networks and for some (in fact for most) initial configurations 
exponential time to converge, no matter which unstable node is switched in each step.
Furthermore, the computational problem: given a network and initial configuration
compute a stable configuration (anyone) that can result from this process
is a PSPACE-complete problem.

Another type of equilibrium problems that can be placed in PLS concerns
finding pure Nash equlibria for games where they are guaranteed to exist.
A (finite) game has a finite set $k$ of players,
each player $i =1,\ldots,k$, has a finite set $S_i$ of
pure strategies and a payoff (utility) function $U_i$
on the product strategy space $S = \Pi_i S_i$; we assume for computational
purposes that $U_i$ takes rational values.
A pure strategy profile $s$ is a member of $S$, i.e. a choice of a pure
strategy $s_i \in S_i$ for each player. It is a pure Nash equilibrium
if no player can improve his payoff by switching unilaterally to another
pure strategy; that is, if $( s_{-i},s'_i)$ denotes the profile where
player $i$ plays strategy $s'_i \in S_i$ and the other players play
the same strategy as in $s$, then $U_i(s) \geq U_i(s_{-i},s'_i)$ for
every $i$ and every $s'_i \in S_i$.
Not every game has a pure Nash equilibrium.
A {\em mixed strategy} for player $i$ is a probability distribution on $S_i$.
Letting $M_i$ denote the set of mixed strategies for player $i$,
the set of mixed strategy profiles is their product $M = \Pi_i M_i$; 
i.e., a mixed strategy profile is a non-negative vector $x$ of length $\sum_i |S_i|$
(i.e. its entries are indexed by all the players' pure strategies)
that is a probability distribution on the set of pure strategies of each player. 
The (expected) payoff $U_i(x)$ of $x$ for player $i$ 
is $\sum x_{1,j_1} \ldots x_{k,j_k} U_i(j_1,\ldots,j_k)$ 
where the sum is over all tuples $(j_1,\ldots,j_k)$ such that
$j_1 \in S_1, \ldots, j_k \in S_k$, and $x_{i,j}$ is the entry of $x$ defining the
probability with which player $i$ plays strategy $j$.
A (mixed) {\em Nash equilibrium} (NE) is a strategy profile $x^*$ such that
no player can increase its payoff by switching to another strategy
unilaterally.  Every finite game has at least one Nash equilibrium \cite{Na}.

For example, a neural network can be viewed as a game with one player for each node,
each player has two pure strategies $+1, -1$ (corresponding to the two states)
and its payoff function has two values 1 (happy) and 0 (unhappy) depending
on its state and that of its adjacent nodes. The stable configurations 
of the network are exactly the pure Nash equilibria of the game.  
This game is a case of a {\em graphical game}:
players correspond to nodes of a graph and the payoff function
of a player depends only on its own strategy and that of its neighbors.
General graphical games may not have pure Nash equilibria.
For an overview of graphical games see \cite{Ke}.

There is a class of games, {\em congestion games}, in which there is always
a pure equilibrium. In a congestion game, there are $k$ players, a finite set $R$ of resources,
the pure strategy set $S_i \subseteq 2^R$ of each player is a family of subsets
of the resources, each resource $r \in R$ has an associated cost function
$d_r: \{0,\ldots,k \} \rightarrow Z$. 
If $s=(s_1,\ldots,s_k)$ is a pure strategy profile,
the congestion $n_r(s)$ of a resource $r$ is the number 
of players whose strategy contains $r$;
the cost (negative payoff) of a player $i$ is $\sum_{r \in s_i} d_r(n_r(s))$.
Rosenthal showed that every congestion game has a pure equilibrium \cite{Ro}.
In fact, the iterative process where in each step, 
if the current pure strategy profile is not
at equilibrium, a player with a suboptimal strategy switches 
to a strategy with a lower cost 
(while other players keep the same strategy) does not repeat any profile and
thus converges in a finite number of steps to an equilibrium.
The proof is by introducing a potential function 
$p(s) = \sum_{r \in R} \sum_{i=1}^{n_r(s)} d_r(i)$ and showing that switching the
strategy of a player to a lower cost strategy results in a reduction of the
potential function by the same amount. Thus, the pure equilibria are
exactly the local optima of the potential function $p(s)$ with respect to
the neighborhood that switches the strategy of a single player.
Computing a pure equilibrium is a local search problem, and it is in PLS 
provided that the costs functions $d_r$ of the resources are polynomial time computable, and
the strategy sets $S_i$ are given explicitly or at least one can determine
efficiently whether a player can improve his strategy for a given profile.
Furthermore, Fabrikant et al. \cite{FPT} showed that the problem is PLS-complete.
They showed that it is complete even in the case of {\em network congestion games}, where
the resources are the edges of a given directed graph, each player $i$ has
an associated source $s_i$ and target node $t_i$ and its set $S_i$ of pure
strategies is the set of $s_i - t_i$ paths; the cost function $d_r$ of each edge $r$
represents the delay as a function of the paths that use the edge, and completeness
holds even for linear delay functions \cite{ARV}. As with other PLS-complete problems,
a consequence of the reductions, which are tight, is that the iterative local improvement
algorithm can take exponential time to converge.
For more information on congestion games see \cite{Vo}.

There are several other games which are in PLS and not known to be in P,
and which are not known (and not believed to be) PLS-complete.
These are not one-shot games, but they are dynamic games played iteratively 
over time (like chess, backgammon etc.).
There are two main types of payoffs for the players in such games: 
in one type, the payoff of a history is an aggregation of rewards obtained
in the individual steps of the history combined via
some aggregation function, such as average reward per step or a discounted sum
of the rewards; in the other type, the payoff obtained depends on the
properties of the history.
We will discuss three such games in this section, and some more in the
following sections.

A {\em simple stochastic game} \cite{Co} is a 2-player game played on a directed graph
$G=(V,E)$ whose nodes represent the positions of the game, and
the edges represent the possible moves. The sinks are labelled
1 or 2 and the nonsink nodes are partitioned into three sets, $V_r$ (random nodes),
$V_1$ (max or player 1 nodes), $V_2$ (min or player 2 nodes); the edges $(u,v)$ out of
each random node $u$ are labelled with probabilities $p_{uv}$ 
(assumed to be rational for computational puposes) that sum to 1. 
Play starts at some initial node (position) and then moves in
each step along the edges of the graph; at a random node the edge is chosen randomly,
at a node of $V_1$ it is chosen by player 1, and at a node of $V_2$ it is chosen
by player 2. If the play reaches a sink labelled 1, then player 1 is the winner,
while if it reaches a sink labelled 2 or it goes on forever, then player 2 is the winner.
The goal of player 1 is to maximize her probability of winning, and
the goal of player 2 is to minimize it (i.e. maximize his own winning probability).
These are zero-sum games (what one player wins the other loses).
For every starting node $s$ there is a well-defined value $x_s$ of the game,
which is the probability that player 1 wins if they both play optimally.
Although the players are allowed to use randomization in each step and have their choice
depend on their entire history, it is known that there are stationary, 
pure (deterministic) optimal strategies for both players. 
Such a strategy $\sigma_i$ for
player $i=1,2$ is simply a choice of an outgoing edge (a successor) for each node in $V_i$,
thus there is a finite number of such pure strategies.
For every pure strategy profile $(\sigma_1,\sigma_2)$ for the two players,
the game reduces to a Markov chain and the values $x_s(\sigma_1,\sigma_2)$
can be computed by solving a linear system of equations.
If the edge probabilities are rational then the optimal values $x_s$ are
also rational, of bit complexity polynomial in the input size. 
If there are only two of the three types of nodes in the graph,
then the optimal strategies and the values $x_s$ can be computed in polynomial time.
For example, if there is no player 2, then the game becomes a Markov decision
process with the goal of maximizing the probability of reaching
a sink labelled 1, which can be optimized by Linear Programming.
When we have all three types of nodes, the decision
problem $x_s \geq 1/2 ?$ (does player 1 win with probability at least 1/2 starting from
position $s$) is in $NP \cap coNP$ (in fact in $UP \cap coUP$), 
and it is a well-known open problem whether it is in P \cite{Co}.
Two (pure) strategies $\sigma_1, \sigma_2$ of the two players form
an {\em equilibrium} if $\sigma_1$ is a best response of player 1 to the strategy $\sigma_2$
of player 2 (i.e. $\sigma_1$ is a maximizing strategy in the Markov
decision process obtained when the strategy of player 2 is fixed to $\sigma_2$),
and vice-versa, $\sigma_2$ is a best response of player 2 to $\sigma_1$.
The equilibria are precisely the optimal strategy pairs.
The problem can be viewed as a local search problem in PLS if we
take the point of view of one player, say player 1: the solution set is the
set of pure strategies of player 1, the value of a strategy $\sigma_1$
is $\sum_{s \in V} x_s(\sigma_1, \sigma_2)$
where $\sigma_2$ is a best response of player 2 to $\sigma_1$, and
the neighbors of $\sigma_1$ are the strategies obtained by switching
the choice of a node in $V_1$. The locally optimal solutions
are the (globally) optimal strategies of player 1.

A {\em mean payoff} game \cite{EM} is a non-stochastic
2-player game played on a directed graph $G=(V,E)$
with no sinks, whose nodes are partitioned into two sets $V_1, V_2$ and whose edges
are labelled by (rational) rewards $r(e), e \in E$. As above, play starts at a node and moves
along the edges, where player 1 chooses the next edge for nodes in $V_1$
and player 2 for nodes in $V_2$ (there are no random nodes here), and play goes on forever.
The payoff to player 1 from player 2 of a history using
the sequence of edges $e_1, e_2, \dots$ is the average reward per step,
$\lim\sup_{n \rightarrow \infty} (\sum_{j=1}^n r(e_j))/n$.
Again there are optimal pure stationary strategies $\sigma_1, \sigma_2$
for the players, and these form a path followed by a cycle $C$; the
payoff (value of the game) is the ratio $\sum_{e \in C} r(e)/|C|$ and is rational
of polynomial bit complexity. 
As shown in \cite{ZP}, the optimal values and optimal strategies
can be computed in pseudopolynomial time (i.e. polynomial time for unary rewards); 
furthermore the problem can be reduced to
simple stochastic games, it is thus in PLS and the decision problem is in $UP \cap coUP$,
but it is open whether it is in P.

A still simpler, nonstochastic 2-player game, called {\em parity game} \cite{EJ}
has been studied extensively in the verification area; it is an important
theoretical question in this area whether this game can be solved in polynomial time.
A parity game is played again on a directed graph $G$ whose nodes are partitioned
into two sets $V_1$, $V_2$ and whose edges are labelled by positive integers.
A history is winning for player 1 (respectively player 2) if the
maximum label that occurs infinitely often in the history is odd (resp. even).
In this game, one of the two players has a pure optimal strategy
that wins on every history that results against every strategy of the other player.
Determining who the winner of the game is (and a winning strategy) reduces to
the decision problem for mean payoff games and in turn to simple stochastic games
\cite{Pu,Jur}.

\section{Fixed Points}

Nash's theorem asserts that every finite game $\Gamma$ has a (generally, mixed)
equilibrium. Nash proved his theorem in \cite{Na} using Brouwer's fixed point theorem:
every continuous function $F$ from a compact convex body to itself has a fixed point,
i.e. a point $x$ such that $x=F(x)$.
Specifically, given a finite game $\Gamma$ with $k$ players $i=1,\ldots,k$,
a finite set $S_i$ of pure strategies and a payoff function $U_i$ for each player,
a mixed strategy profile is a vector $x=(x_{ij}| i=1,\dots,k; j =1,\ldots,|S_i|)$, 
which lies on the  product $\Delta$ of the $k$ unit simplexes 
$\Delta_i= \{ y \in R ^{|S_i|} | \sum_{j=1}^{|S_i|} y_j =1; y \geq 0 \}$.
Nash defined the following function from $\Delta$ to itself:
$ F_\Gamma(x)_{(i,j)} \doteq \frac{x_{i,j} +  \max \{ 0,  g_{i,j}(x)\}}
{1 + \sum^{|S_i|}_{l=1}  \max \{ 0, g_{i,l}(x)\}}$,
where $g_{i,j}(x)$ is the (positive or negative) ``gain'' in payoff of player $i$ 
if he switches to pure strategy $j$ while the other players continue to play according
to $x$; $g_{i,j}(x)$ is a (multivariate) polynomial in $x$.
Nash showed that the fixed points of $F_\Gamma$ are precisely the equilibria 
of the game $\Gamma$. There are several alternative proofs
of Nash's theorem, all using Brouwer's theorem (with different functions $F$)
or the related Kakutani's theorem (for fixed points of multivalued maps). 
Note that the underlying solution space here, 
$\Delta$, is continuous, not discrete and finite. 
Furthermore, even if the payoff functions of the game are rational-valued,
for 3 or more players it may be the case that all equilibria are irrational.

Market equilibria is another important application of fixed point
theorems.  Consider the following exchange model \cite{Sc2}.  We have
$m$ agents and $n$ commodities.  The agents come to the market with an
initial supply of commodities, which they exchange for their prefered
ones; each agent sells his supply at the prevailing prices, and buys
his preferred bundle of commodities.  For each vector $p$ of prices
for the commodities, each agent $\ell$ has an (positive or negative)
`excess demand' (=demand-supply) $g^\ell_i(p)$ for each commodity $i$.
Standard assumptions are that the functions $g^\ell_i(p)$ (i) are
homogeneous of degree 0, thus the price vectors may be normalized to
lie on the unit simplex $\Delta_n$, (ii) they satisfy Walras' law
$\sum_{i=1}^n p_i g^\ell_i(p) = 0$, (iii) they are continuous on the unit
simplex.  Let $g_i(p) = \sum_\ell g^\ell_i(p)$ be the (total) market excess
demand for each commodity $i$.  The functions $g_i(p)$ satisfy the
same constraints.  A vector $p$ of prices is an {\em equilibrium} if
$g_i(p) \leq 0$ for all $i$ (demand does not exceed supply), with
equality for all commodities $i$ that have $p_i >0$.  Brouwer's
theorem can be used to show the existence of equilibria.  Namely, the
equilibria are the fixed points of the function $F: \Delta_n \mapsto
\Delta_n$, defined by the formula $F_i(p) = \frac{p_i +
\max(0,g_i(p))}{1+\sum_{j=1}^n \max(0,g_j(p))}$.  In fact, the
equilibrium existence theorem can be conversely used to show Brouwer's
theorem: from a Brouwer function one can construct an economy whose
equilibria correspond to the fixed points of the function \cite{Uz}.
In the classical Arrow-Debreu market model \cite{AD}, the user
preferences for the commodities are modeled by utility functions,
which in turn induce the excess demand functions (or correspondences,
i.e. multivalued maps), and more generally the model includes also
production.  Under suitable conditions, the existence of equilibria is
derived again using a fixed point theorem (Kakutani in \cite{AD}, or
Brouwer in alternative proofs \cite{Ge}).  As shown in a line of work
by Sonnenschein, Mantel, Debreu and others (see e.g. \cite{De}),
essentially any function satisfying the standard conditions can arise
as the excess demand function in a market for suitably defined utility
functions for the users .  Thus, there is a tight connection between
fixed points of general functions and market equilibria.

A number of other problems from various domains can be cast as fixed
point computation problems, i.e., every instance $I$ of a problem
is associated with a function $F_I$ over some domain so that
the sought objects $Ans(I)$ are fixed points of $F_I$;
in some cases, we may only want a specific fixed point of the function.
We will mention several more examples in this section.
Recall the simple stochastic game from the last section. 
The vector $x= (x_s | s \in V)$ of winning probabilities for Player 1
satisfies the following system of equations $x=F(x)$, with
one equation for each node $s$:
if $s$ is a sink labelled 1 (respectively 2) then $x_s =1$ (resp. $x_s =0$);
if $s \in V_r$ then $x_s = \sum_{(s,v) \in E} p_{sv} x_v$;
if $s \in V_1$ then $x_s = \max \{x_v | (s,v) \in E \}$;
if $s \in V_2$ then $x_s = \min \{x_v | (s,v) \in E \}$.
In general there may be multiple solutions, however the
system can be preprocessed so that there is a unique solution
in the unit cube $C_n = \{x | 0 \leq x_s \leq 1, \forall s \in V \}$.

Stochastic games were originally introduced by Shapley in \cite{Sh}
in a more general form, where players can move simultaneously.
As shown in \cite{Co}, simple stochastic games can be reduced to Shapley's game.
In {\em Shapley's game} there is a finite set $V$ of states, each state $u$
has an associated one-shot zero-sum finite game with a reward (payoff) matrix $A_u$
whose rows (resp. columns) correspond to the actions (pure strategies) of Player 1 (resp. 2).
If the play is in state $u$ and the players
choose actions $i,j$ then Player 1 receives
reward $A_u[i,j]$ from Player 2, the game stops
with probability $q_{ij}^u >0$, and it transitions to state $v$ with
probability $p_{ij}^{uv}$, where $q_{ij}^u + \sum_v p_{ij}^{uv}=1$.
Since there is at least positive probability $q= \min\{ q_{ij}^u | u,i,j \} >0$
of stopping in each step, the game stops a.s. in a finite number of steps.
(Another standard equivalent formulation is as a discounted game,
where the game does not stop but future rewards are discounted 
by a factor $1-q$ per step).  
The goal of Player 1 is to maximize 
(and of  Player 2 to minimize) the total expected reward, 
which is the value of the game.
We want to compute the vector  $x=(x_u | u \in V)$
of game values for the different starting states $u$.  
As usual all rewards and probabilities
are assumed to be rationals for computational purposes.
The values in general may be irrational now however.
The vector $x$ satisfies a fixed point set of equations $x=F(x)$,
as follows. For each state $u$, let $B_u(x)$ be the matrix, 
indexed by the actions of the players,
whose $i,j$ entry is $A_u[i,j] + \sum_v p_{ij}^{uv} x_v$,
and let $Val(B_u(x))$ be the value of the one-shot zero-sum game
with payoff matrix $B_u(x)$.
Then $x=F(x)$ where $F_u(x)=Val(B_u(x)),~u\in V$.
The function $F$ is a Banach function (a contraction map) under the $L_{\infty}$
norm with contraction factor $1-q$, and thus it has a unique
fixed point, the vector of values of the game. 

{\em Branching processes} are a basic model of stochastic evolution,
introduced first in the single type case by Galton and Watson in the 19th century to
study population dynamics, and extended later to the  multitype case
by Kolmogorov and Sevastyanov, motivated by biology.
A branching process has a finite set $T$ of $n$ types, for each type $i \in T$
there is a finite set of `reproduction' rules of the form
$i \stackrel{p_{ij}}{\rightarrow}  v_{ij}, j=1,\ldots,m_i$,
where $p_{ij} \in [0,1]$ is the probability of the
rule (thus, $\sum_{j=1}^{m_i} p_{ij}=1)$ and  $v_{ij} \in \nat^n$ is a vector
whose components specify the number of offsprings of each type
that an entity of type $i$ produces in the next generation.
Starting from an initial population, the process evolves from one generation
to the next according to the probabilistic reporuction rules. The
basic quantity of interest is the probability $x_i$ of extinction of
each type: the probability that if we start with one individual of
type $i$, the process will eventually die.
These can be used  to compute the extinction probability for  any initial
population and are the basic for more detailed statistics of the process.
As usual, we assume that the probabilities of the rules are rational.
However the extinction probabilities are in general irrational.
The vector $x$ satisfies a set of fixed point equations $x=F(x)$,
where $F_i(x)$ is the polynomial 
$\sum_{j=1}^{m_i} p_{ij} \Pi_{k=1}^n (x_k)^{v_{ij}[k]}$.
Note that $F_i(x)$ has positive coefficients, thus $F$ is a monotone
operator on $\real^n_{\geq 0}$ and thus has a Least Fixed Point (LFP);
the LFP is precisely the vector of extinction probabilities of
the  branching process. 
For more information on the theory of branching processes and their applications
see \cite{Ha,HJV}.

{\em Stochastic context-free grammars} (SCFG) are context-free grammars where the
production rules have associated probabilities. They have been studied extensively
in Natural Language Processing where they are an important model \cite{MS},
and have been used also in biological sequence analysis.
A basic quantity of interest is the probability of the language generated
by a SCFG; again this may be an irrational number even if all
the probabilities of the production rules are rational.
The analysis of SCFG's is closely related to that of branching processes.

A model that encompasses and generalizes both of branching processes
and SCFG's in a certain precise sense,
is the {\em Recursive Markov chains} (RMC) model \cite{EY1} and the equivalent model of
{\em Probabilistic Pushdown machines} \cite{EKM}. Informally, a RMC is
a collection of Markov chains that can call each other in a potentially
recursive manner like recursive procedures. The basic quantities of interest
are the termination probabilities. These probabilities obey again a system
of fixed point equations $x=F(x)$, where $F$ is a vector of polynomials
with positive coefficients; the least fixed point of the system 
gives the termination probabilities of the RMC.
Generalization to a setting where the dynamics are not completely
probabilistic but can be controlled by one or more players
leads to {\em recursive Markov decision processes and games} \cite{EY2,EY3,EY5}.
For example, we may have a branching process, where 
the reproduction can be  influenced by
players who want to bias the
process towards extinction or survival.
This results in fixed point systems of equations 
involving monotone polynomials and the min and max
operators. 

All of the above problems are total (single-valued or multi-valued) search problems,
in which the underlying solution space is continuous.
In all of these problems we would ideally like to
compute exactly the quantities of interest if possible (if they are rational),
and otherwise, we would like to bound them and answer decision questions
about them (eg. is the value of a stochastic game $\geq 1/2$?, does
a RMC terminate with probability 1?) or to approximate them within desired
precision, i.e. compute a solution $x$ that is within $\epsilon$ of
an/the answer $x^*$ to the search problem (eg., approximate within additive
error $\epsilon$ the extinction probabilities of a branching process,
or compute a mixed strategy profile for a game that is within $\epsilon$
of a Nash equilibrium). In the approximation problem we would like
ideally polynomial time in the size of the input and in $\log(1/\epsilon)$ 
(the number of bits of precision). We refer to the approximation of
an answer to a search problem as above
as {\em strong} approximation (or the `near' problem) to
distinguish it from another notion of approximation,
which we call the {\em weak} approximation (or the `almost' problem)
that is specific to a fixed point formulation of a search problem
via a function $F$:
a weak $\epsilon$-approximation is a point $x$ such that 
$|x-F(x)| \leq \epsilon$ (say in the $L_{\infty}$ norm).
Note that a search problem may be expressible in different ways
as a fixed point problem using different functions $F$,
and the notion of weak approximation may depend on the function that
is used; the strong approximation notion is intrinsic to the search problem
itself (does not depend on $F$). 
For many common fixed point problems
(formally, for polynomially continuous functions \cite{EY4}),
including all of the above problems, 
weak approximation reduces to strong,
i.e., given instance $I$ and (rational) $\epsilon >0$, 
we can define a (rational) $\delta >0$ of bit-size polynomial
in that of $\epsilon$ and in $|I|$ such that every (strong) 
$\delta$-approximation $x$ to an answer to the search problem
(i.e., approximation to a fixed point of the function $F_I$ corresponding
to the instance $I$) is a weak $\epsilon$-approximate fixed point
(i.e., satisfies $|x-F_I(x)| \leq \epsilon$).
The converse relation does not hold in general; in particular, it does not hold
for Nash equilibria and the Nash function $F_{\Gamma}$.

We discuss briefly now algorithms for such fixed point problems.
For a Banach function $F_I$ we can start at any point $x_0$, and
apply repeatedly $F_I$. The process will converge to the unique fixed
point. If the contraction factor $1-q$ is a constant $<1 $, then convergence is polynomial,
but if the margin $q$ from 1 is very small, inverse exponential in the size of the input
$I$ (as is generally the case, for example in Shapley's game),
then convergence is  slow. 

For a monotone function $F_I$ for which we want to compute the least fixed point, 
as in many
of the examples above (stochastic games, branching processes, RMC etc.), 
we can start from $x_0=0$ (which is lower than the LFP) and apply repeatedly $F_I$;
the process will converge to the desired LFP, but again convergence is generally slow.
Note that for many of these problems, obtaining a weak $\epsilon$-approximation
for $\epsilon$ constant or even inverse polynomial, $|I|^{-c}$ is easy: 
for  example, in a simple stochastic game or a branching process, the vector $x$
is bounded from above by the all-1 vector and $F_I^k(0), k=0,1,2,\ldots$ increases  
monotonically with $k$, so after at most $n/\epsilon$ iterations we will
get a weak $\epsilon$-approximate fixed point $x$. However, such a point $x$ is of  
no use in estimating the actual values or probabilities that we want to compute.
Approximating the value of a simple stochastic game even within additive error 1/2
is an open problem.

For general Brouwer functions $F$ we cannot simply apply iteratively $F$
from some starting point $x_0$ and hope to converge to a fixed point.
There is extensive algorithmic work on the approximate computation of
Brouwer fixed points, starting with Scarf's fundamental algorithm \cite{Sc1},
The standard proof of Brouwer's theorem involves a combinatorial lemma,
Sperner's lemma, combined with a (generally nonconstructive) compactness argument.
Scarf's algorithm solves constructively Sperner's problem, and computes
a weak $\epsilon$-approximate fixed point for the function. 
Briefly, it works as follows. Assume wlog that the domain is the unit simplex
$\Delta_n =\{ x \geq 0 | \sum_i x_i=1 \}$, and consider a simplicial
subdivision of $\Delta_n$ into
simplices of sufficiently small diameter $\delta$, so that
$|x-y| \leq \delta$ implies $|F(x) - F(y)| \leq \epsilon /n$.
Label (``color'') each vertex $v$ of the subdivision by an index $i=1,\ldots,n$
such that $v_i > F_i(v)$; if $v$ is not a fixed point there is
at least one such index, if $v$ is a fixed point then
label $v$ with say $\arg_i \max(v_i)$. Note that the unit vectors $e_i$
at the $n$ corners of the simplex $\Delta_n$ are labelled $i$,
and all vertices on the facet $x_j=0$ are labelled with an index $\neq j$.
Sperner's lemma implies then that the subdivision
has at least one panchromatic simplex, i.e. a small simplex $S$ 
whose vertices have distinct labels. From the definition of the labels
and the choice of $\delta$ it follows that any point $x \in S$ satisfies
$|F(x) - x| \leq \epsilon$. Scarf's algorithm starts with a suitable
subdivision and a boundary simplex whose vertices have $n-1$ distinct indices (all except one),
and then keeps moving to an adjacent simplex through the face with the $n-1$ indices;
the process cannot repeat any simplex of the subdivision, so it will
end up at a panchromatic simplex $S$. Note that $S$ may not contain any
actual fixed points, and in fact may be located far from all of them, but
any point $x$ of $S$ is a weak $\epsilon$-approximation.
If we take finer and finer subdivisions letting the diameter $\delta$ go down
to 0, then the resulting sequence of weakly approximate fixed points
must contain (by compactness) a subsequence that converges to a point,
which must be a fixed point; this latter part however is nonconstructive
in general.

There are several other subsequent methods for computing (approximate) fixed points,
e.g. Newton-based, and homotopy methods (some of these assume differentiability and
use also the derivatives of the function).
Scarf's algorithm, as well other general-purpose algorithms, treat the function $F$
as a black box. Such black box algorithms must take exponential time in the worst case
to compute a weak approximation \cite{HPV}. Furthermore, for strong approximation 
no finite amount of time is enough in the black box model \cite{Si}, 
and there are also noncomputability results for computing equilibria and
fixed points for a model where the function is given via a Turing machine \cite{Ko,RW}.
However, the restriction to black box access is a severe one, and
the results do not mean that any of the specific problems we want to solve 
(for example, Nash equilibria) is necessarily hard.

\section{Rational equilibria, Piecewise Linear Functions and the Class PPAD}

Consider a 2-player finite game, with the payoffs given
explicitly in terms of the two payoff matrices $A_1, A_2$ of
the two players (i.e., the game is presented in {\em normal form}). 
Computing a specific Nash equilibrium, such as one that maximizes the
payoff to one of the players, or to all the players, is NP-hard \cite{GZ}.
However, the search problem that asks for {\em any} Nash equilibrium
is a different, `easier' problem, and is unlikely to be NP-hard.  

The 2-player case of the Nash equilibrium problem can be viewed either as a continuous or
as a discrete problem, like Linear Programming: We can consider LP either 
as having a continuous solution space, namely all the real-valued points in the
feasible polyhedron, or as having a discrete solution space, namely the vertices of
the polyhedron or the feasible bases. Similarly, for 2-player games which correspond
to a Linear Complementarity problem.
A mixed strategy profile is a Nash equilibrium iff every pure strategy of
each player is either at 0 level (not in the support) or is
a best response to the strategy of the other player.
Assumming the game is nondegenerate (we can always ensure this by a small perturbation)
the supports of the mixed strategies determine uniquely the equilibrium:
we can set up and solve a linear system of equations which equate the
payoffs of the pure strategies in the support of each player,
and check that the solution satisfies the appropriate inequalities
for the pure strategies that are not in the supports. 
One  consequence of this is that if the payoffs are rational
then there are rational equilibria, of polynomial bit complexity in the
input size, and they can be computed exactly.
A second consequence is that Nash's theorem in this case can be proved
directly, without resorting to a fixed point theorem,
and algorithmically, namely by the Lemke-Howson algorithm \cite{LH}.
The algorithm has similar flavor to Scarf's algorithm for fixed points.
Mixed profiles can be labelled (`colored') by the set of pure strategies
that are not in the support or that are best responses to the other
player's strategy.
The equilibria are the mixed profiles that are panchromatic, i.e.,
labeled with all the pure strategies of both players.
Briefly, the algorithm starts from an artificial point
that has all the colors except one, and then follows a path through
a sequence of LP-like pivots, until it arrives at a panchromatic point
(profile), 
which must be an equilibrium; the algorithm cannot repeat any point,
because at any point there
are only two possible pivots, one forward and one backward,
and there is a finite number of points (supports) so it terminates. 
It is known that the algorithm takes exponential time in the worst case \cite{SS}.

Papadimitriou defined in \cite{Pa} a complexity class, PPAD, that
captures the basic principles of these path-following algorithms:
There is a finite number of candidate solutions, and an underlying 
directed graph of moves between the solutions where
each solution has at most one forward and one backward move,
i.e., the graph consists of a set of directed paths, cycles
and isolated nodes;  a source of one path is an artificial starting solution, 
and every other endpoint (source or sink) of every path is an answer to the problem
(eg., an equilibrium). Formally, a search problem $\Pi$ is in PPAD if 
each instance $I$ has a set $S(I)$ of solutions 
which are (strings) polynomially bounded in the input size $|I|$, 
and there are polynomial-time algorithms for the following tasks:
(a) test whether a given string $I$ is an instance of $\Pi$ and 
if so compute a initial solution $s_0$ in $S(I)$, (b) given $I,s$, 
test whether $s \in S(I)$
and if so compute a successor $succ_I(s) \in S(I)$ and
a predecessor $pred_I(s) \in S(I)$, such that $pred_I(s_0) = s_0$,
$succ_I(s_0) \neq s_0$, and $pred_I(succ_I(s_0)) = s_0$.
The $pred$ and $succ$ functions induce a directed graph $G=(S(I),E)$,
where $E= \{ (u,v) | u \neq v, succ_I(u)=v, pred_I(v)=u \}$,
and the answer set to the instance $I$ of the
search problem, $Ans(I)$, is the set of nodes of $G$, other  than $s_0$
that have indegree + outdegree = 1, i.e., are endpoints of the paths;
note that $Ans(I) \neq \emptyset$ because there must be at least one more endpoint
besides $s_0$.
As is customary, the class is closed under polynomial-time reduction,
i.e., if a search problem $\Pi'$ reduces to a problem $\Pi$ that
satisfies the above definition, then $\Pi'$ is considered also to
belong to PPAD.
Papadimitriou defined two other variants of this class in \cite{Pa}, PPA in which the
underlying graph is undirected, and PPADS in which the  graph is directed
and the answer set consists only of the sinks of the paths.
However, PPAD is the more interesting and richer of these classes in terms
of natural problems.

The class PPAD lies somewhere between P and TFNP: all search problems in PPAD are total,
and furthermore, for a given instance $I$, we can guess a solution $s$ and verify
that it is an answer. Thus, as in the case of PLS, problems in PPAD cannot be
NP-hard unless NP=coNP.

By virtue of the Lemke-Howson algorithm, the Nash equilibrium problem for
2-player (normal form) games is in PPAD. For 3 or more players we cannot say that
the Nash problem is in PPAD; for one thing the equilibria are irrational.
But the following approximate $\epsilon$-Nash version is in PPAD \cite{DGP}.
An $\epsilon$-Nash equilibrium of a game is a (mixed) strategy profile
such that no player can improve its payoff by more than $\epsilon$
by switching unilaterally to another strategy. (Note, this is not
the same as being $\epsilon$-close to a Nash equilibrium.)
The $\epsilon$-Nash problem is: given a normal form game $\Gamma$ (with rational
payoffs) and a rational $\epsilon >0$, compute an $\epsilon$-Nash equilibrium of
$\Gamma$. (Note that $\epsilon$ is given as usual in binary, 
so polynomial time means polynomial in $|\Gamma|$ and $log(1/\epsilon)$.)
The complexity of the Nash problem was one of the main motivations 
for the original introduction of PPAD. A recent sequence of papers
culminated in showing that the Nash equilibrium problem for 2-player games
is PPAD-complete \cite{DGP,CD1}, that is, if the problem
can be solved in polynomial time, then so can all the problems in PPAD. 
Furthermore, even the $\epsilon$-Nash equilibrium problem for $\epsilon$ 
inverse polynomial, i.e. even with $\epsilon$ given in unary, 
is also PPAD-complete for 2-player games \cite{CDT}.
For all constant $\epsilon$, an $\epsilon$-Nash equilibrium can be computed in 
quasipolynomial time \cite{LMM}.

Another basic PPAD-complete problem is (a formalization of) the Sperner problem.
The 2D case concerns the unit simplex (triangle) $\Delta_3$
and its simplicial subdivision (i.e., triangulation) with vertices $v=(i_1/n,i_2/n,i_3/n)$
with $i_1+i_2+i_3=n$. The 2D Sperner problem is as follows. 
The input consists of a number $n$ in binary and a Boolean circuit which
takes as input three natural numbers $i_1, i_2, i_3$ with $i_1+i_2+i_3=n$
and outputs a color $c \in \{1,2,3\}$, with the restriction
that $i_c \neq 0$. The problem is to find a trichromatic triangle,
i.e. three vertices (triples) with pairwise distances $1/n$ that have distinct
colors. The problem is in PPAD by Scarf's algorithm. 
The 3D Sperner problem was shown PPAD-complete in \cite{Pa},
and the 2D case was shown complete in \cite{CD2}.

The original paper showed PPAD-completeness also for a discretized version
of Brouwer and related theorems (Kakutani, Borsuk-Ulam) in the
style of the Sperner problem: 
a Brouwer function in 3D is given in terms of a binary number $n$
(the resolution of a regular grid subdivision of the unit 3-cube) 
and a Boolean circuit that takes as input three natural numbers
$i_1,i_2,i_3$ between 0 and $n$ and outputs the value 
of the function at the point $(i_1/n,i_2/n,i_3/n)$.
The function is then linearly interpolated in the rest of the unit
cube according to a standard simplicial subdivision with the grid points
as vertices.
The paper showed also completeness for a discretized version
of the market equilibrium problem for an exchange economy.

In \cite{CSVY} Codenoti et al. show PPAD-completeness of the price equilibrium 
for a restricted case of Leontief exchange economies,
i.e. economies in which each agent $i$ wants commodities in
proportion to a specified (nonnegative) vector $(a_{i1},\ldots,a_{ik})$;
that is, the utility function of agent $i$
is $u_i(x) = \min\{x_{ij}/a_{ij}| j=1,\ldots,k; a_{ij} \neq 0\}$.
In general, such economies may not have an equilibrium,
and it is NP-hard to determine if there is one \cite{CSVY}.
However, a restricted subclass of Leontief economies
has equilibria and is equivalent to the Nash equilibrium problem for 2-player games.
This restricted Leontief class is as follows:
the agents are partitioned into two groups,
every agent brings one distinct commodity to the market,
and agents in the first group want commodities only of agents in the
second group and vice-versa.

The class PPAD cannot capture of course general Brouwer functions since
many of them have irrational fixed points as we saw in the last section. 
(We could discretize such a function, but then the resulting approximating
function has new fixed points, which may have no relation
and can be very far from the fixed points of the original function.)
However there is a natural class of functions that are
guaranteed to have rational fixed points, which are in PPAD
and in a sense characterize the class \cite{EY4}. 
Consider the search problem $\Pi$ of computing a fixed point
for a family of Brouwer functions 
${\mathcal F}= \{F_I | I$ an instance of $\Pi \}$.
We say that $\Pi$ is a {\em polynomial piecewise linear} problem if
the following hold: 
For each instance $I$, the domain is divided by hyperplanes into polyhedral cells, 
the function $F_I$ is linear in each cell and is of course continuous over the
whole domain. The coefficients of the
function in each cell and of the dividing hyperplanes
are rationals of size bounded by a polynomial in $|I|$.
These are not given explicitly in the input, in fact there may be
exponentially many dividing hyperplanes and cells.
Rather, there is an oracle algorithm that runs in time
polynomial in $|I|$ which generates a sequence of
queries of the form $ax \leq b?$ adaptively 
(i.e., the next query depends on
$I$ and the sequence of previous answers), and at the end
either outputs `No' (i.e., $x$ is not in the domain) 
or identifies the cell of $x$ and outputs the coefficients $c,c'$ 
of the function $F_I(x) =cx+c'$.
As shown in \cite{EY4}, all polynomial piecewise linear problems 
are in PPAD (they all have rational fixed points of polynomial size).
Examples include the simple stochastic games, 
the discretized Brouwer functions obtained from linear interpolation
on a grid, and the Nash equilibrium problem for 2-player games
(Nash's function in nonlinear even for 2 players, but there is another
piecewise linear function whose fixed points are also exactly
the Nash equilibria).

The class PPAD captures also the approximation in the weak (`almost') sense
for a broad class of Brouwer functions,
and in some cases also the strong approximation (`near') problem \cite{EY4}.
Consider a family of functions ${\mathcal{F}}= \{F_I\}$.
We say ${\mathcal F}$ is {\em polynomially computable}
if for every instance $I$ and rational vector $x$ in the domain, 
the image $F_I(x)$ is rational and can be computed in time
polynomial in the size of $I$ and of $x$. 
${\mathcal F}$ is called 
{\em polynomially continuous} if 
there is a polynomial $q(z_1,z_2)$ such that for
all instances $I$ and all rational
$\epsilon >0$, there is a rational $\delta > 0$ such that $size(\delta) 
\leq q(|I|,size(\epsilon))$ and such that for all $x, y \in D_I$,
$| x - y | < \delta \  \Rightarrow \  |F_I(x)-F_I(y)| < \epsilon$.
If ${\mathcal F}$ is polynomially computable and polynomially continuous,
then the weak approximation problem (given instance $I$ and
rational $\epsilon > 0$, compute a weakly $\epsilon$-approximate fixed point
of $F_I$) is in PPAD by virtue of Scarf's algorithm. 
Furthermore, if the functions $F_I$
happen to be also contracting with contraction rate $< 1- 2^{-poly(|I|)}$,
then strong approximation reduces to weak approximation,
and the strong approximation problem (given $I, \epsilon$, compute
a point $x$ that is within $\epsilon$ of some
fixed point $x^*$ of $F_I$) is also in PPAD; Shapley's problem
is an example that satisfies this condition.
Moreover, if in addition the functions $F_I$ have rational
fixed points of polynomial size, then strongly $\epsilon$-approximate fixed points
with small enough $\epsilon$ can be rounded to get exact fixed points, 
and thus the exact problem is in PPAD; simple stochastic games, 
perturbed with a small discount \cite{Co}, are such an example.

\section{Irrational Equilibria, Nonlinear Functions, and the Class FIXP}

Games with 3 or more players are quite different from 2-player games: 
Nash equilibria are generally irrational; knowing the support of an equilibrium
does not help us much, and there may be many different such equilibria.
There are many search problems as we saw in Section 3, and in particular
many problems that can be cast in a fixed point framework, where the
objects that we want to compute (the answers) are irrational.
Of course we cannot compute them exactly in the usual Turing machine
model of computation. One can consider the exact computation and 
the complexity of such search problems in a real model of computation \cite{BCSS}.
In the usual (discrete) Turing model of computation and complexity,
we have to state carefully and precisely what is the (finite) information
about the solution that we want to compute, as the nature
of the desired information can actually affect the complexity of the problem,
i.e., some things may be easier to compute than others.
That is, from a search problem $\Pi$ with a continuous solution space,
another search problem $\Pi'$ is derived with a discrete space. 
Several types of information are potentially of interest, leading to
different problems $\Pi'$.

Consider for example Shapley's stochastic game. Some relevant questions
about the value of the game are the following: 
(i) {\em Decision problem}: Given game $\Gamma$ and rational $r$, is the value of
the game $\geq r$?, (ii) {\em Partial computation}: Given $\Gamma$, integer $k$,
compute the $k$ most significant bits of the value, (iii) {\em Approximation}: 
Given $\Gamma$, rational $\epsilon>0$, compute an $\epsilon$ approximation to 
the value. Similar questions can be posed about the optimal strategies
of the players. The value of a game is a problem with a unique answer; for 
multivalued search problems (e.g., optimal strategy, Nash equilibrium etc.) care
must be taken in the statement of the discrete problems (e.g., the decision problem) 
so that it does not become harder than the search problem itself;
in general, the requirement in the multivalued case is that
the response returned for the discrete problem should be valid for some answer to 
the continuous search problem. 
As we said in the previous section, the approximation problem for the
value of Shapley's game is in PPAD (and it is open whether it is P).
The decision (and partial computation) problem however seems to be
harder and it is not at all clear that it is even in NP; 
in fact showing that it is in NP would answer a well-known longstanding
open problem. The same applies to many other problems.
The best upper bound we know for the decision (and partial computation)
problem for Shapley's games and for many of the other fixed point problems
listed in Section 3 (eg., branching processes, RMCs etc) is PSPACE.

The {\em Square Root Sum} problem (\sqrtsum{} for short) is the following  problem:
given positive integers $d_1,\ldots,d_n $ and $k $, decide
whether $\sum^n_{i=1} \sqrt{d_i} \leq k$.
This problem arises often for example in geometric computations,
where the square root sum represents the sum of Euclidean distances
between given pairs of points with integer (or rational)
coordinates; for example, determining whether the length of
a specific spanning tree, or a TSP tour of given points on the
plane is bounded by a given threshold $k$
amounts to answering such a problem.
This problem is solvable in PSPACE, but it has been a major
open problem since the 1970's 
(see, e.g., \cite{GGJ,Ti}) 
whether it is solvable even in NP (or better yet, in P).  
A related, and in a sense more powerful and fundamental, problem is
the {\em \posSLP{} problem}: given a division-free straight-line program,
or equivalently, an arithmetic circuit with operations $+, -, *$ and
inputs 0 and 1, and a designated output gate,
determine whether the integer $N$ that is the output of the circuit is positive.
The importance of this problem was highlighted in \cite{Al+},
which showed that it is the key problem in understanding the
computational power of the Blum-Shub-Smale model of real computation \cite{BCSS} 
using rational numbers as constants,
in which all operations on rationals take unit time, no matter their size;
importantly, integer division (the floor function) is not allowed
(unit cost models with integer division or logical bit operations
can solve in polynomial time all PSPACE problems, 
see e.g. \cite{EB} for an overview of machine
models and references).
This is a powerful model in which the \sqrtsum{} problem 
can be decided in polynomial time
\cite{Ti}). Allender et al. \cite{Al+} showed that the set of 
discrete decision problems that 
can be solved in P-time in this model is equal to
$P^{\posSLP{}}$, i.e. problems solvable in P using a subroutine for \posSLP{}. 
They showed also that \posSLP{} and \sqrtsum{} 
lie in the Counting Hierarchy (a hierarchy above PP).

The \sqrtsum{} problem can be reduced to the decision version of
many problems: the Shapley problem \cite{EY4}, 
concurrent reachability games \cite{EY3}, 
branching processes, Recursive Markov chains \cite{EY1},
Nash equilibria for 3 or more players \cite{EY4}.
The \posSLP{} problem reduces also to several of these.
Hence placing any of these problems in NP would
imply the same for \sqrtsum{} and/or \posSLP{}.
Furthermore, for several problems, the approximation of
the desired objects is also at least as hard.
In particular, approximating the termination probability of a
Recursive Markov chain within any constant additive error $< 1$
is at least as hard as the \sqrtsum{} and the \posSLP{} problems \cite{EY4}.

A similar result holds for the approximation of Nash equilibria
in games with 3 or more players. Suppose we want to
estimate the probability with which
a particular pure strategy, say strategy 1 of player 1, is played in
a Nash equilibrium (any one); obviously, the value 1/2 
estimates it trivially with error $\leq 1/2$.
Guaranteeing a constant error $< 1/2$ is at least as hard
as the \sqrtsum{} and the \posSLP{} problems \cite{EY4},
i.e. it is hard to tell whether the strategy will be played
with probability very close to 0 or 1.

The constructions illustrate also the difference between strong and weak
approximate fixed points generally, and for specific problems in particular.
Recall that for RMCs we can compute very easily a weak $\epsilon$-approximate
fixed point for any constant $\epsilon >0$; however it is apparently much harder
to obtain a strong approximation, i.e. approximate the actual probabilities
within any nontrivial constant.
In the RMC case the weak approximation is irrelevant.
However, in the case of Nash equilibria, the weak approximation of Nash's
function is also very natural and meaningful:
it is essentially equivalent to the notion of $\epsilon$-Nash equilibrium
(there is a small polynomial change in $\epsilon$ in each direction).
For every game $\Gamma$ and $\epsilon >0$, we can choose a $\delta$
of bit-size polynomial in the size of $\Gamma$ and $\epsilon$ so that
every strategy profile that is within distance $\delta$ of a Nash equilibrium
is $\epsilon$-Nash (i.e. all strongly approximate points are 
also weakly approximate with a `small' change in $\epsilon$).
However, the converse is not true:
For every $n$ there is a 3-player game of size $O(n)$, with 
an $\epsilon$-Nash equilibrium, $x'$, where $\epsilon = 1/2^{2^{\Omega(n)}}$,
such that $x'$ has distance $1- 2^{-poly}$ (i.e., almost 1) 
from every Nash equilibrium \cite{EY4}.

For 2-player games, as we said there is a direct, algorithmic proof
of the existence of Nash equilibria (by the Lemke-Howson algorithm).
But for 3 and more players, the only proofs known are through a fixpoint theorem
(and there are several proofs known using different Brouwer functions
or Kakutani's theorem).
In \cite{EY4} we defined a class of search problems, FIXP, that can be
cast as fixed point problems of functions that use the usual algebraic operations
and max, min, like Nash's function, and the other functions for
the problems discussed in Section 3. 
Specifically, FIXP is the class of  search problems
$\Pi$, such that there is a polynomial-time algorithm which, given an instance
$I$, constructs an algebraic circuit (straight-line program) $C_I$ over the
basis $\{+,*,-,/,\max,\min \}$, with rational constants,
that defines a continuous function $F_I$ from
a domain to itself (for simplicity, standardized to be the unit cube,
other domains can be embedded into it), with the property that
$Ans_{\Pi}(I)$ is the set of fixed points of $F_I$.
The class is closed as usual under reductions.
In the usual case of discrete search problems,
a reduction from problem $A$ to problem $B$ consists of two
polynomial-time computable functions, a function $f$ that maps instances $I$ of
$A$ to instances $f(I)$ of $B$, and a second function $g$ that maps
solutions $y$ of the instance $f(I)$ of $B$ to solutions $x$ of the instance $I$ of $A$.
The difference here is that the solutions are real-valued, not discrete,
so we have to specify what kind of functions $g$ are allowed. 
It is sufficient to restrict the reverse function $g$
to have a particularly simple form:
a separable linear transformation with polynomial-time computable rational
coefficients; that is, $x=g(y)$, where each $g_i(y)$ is of the form
$a_i y_j +b_i$ for some $j$, where $a_i, b_i$ are rationals computable from  $I$
in polynomial time.
Examples of problems in FIXP include: Nash equilibrium for normal form
games with any number of players, price equilibrium in
exchange economies with excess demand functions given by
algebraic formulas or circuits, the value (and optimal strategies) 
for Shapley's stochastic games, 
extinction probabilitites of branching processes, and
probability of languages generated by stochastic context-free grammars.

FIXP is a class of search problems with continuous solution spaces,
and corresponding to each such problem $\Pi$, there are the
associated discrete problems: decision, approximation etc.
All the accociated discrete problems can be expressed in the existential
theory of the reals, and thus, using decision procedures for this theory \cite{Ca,Re},
it follows that they are all in PSPACE.
As we mentioned, many of these problems are at least as hard
as the \sqrtsum{} and the \posSLP{} problems, for which the current
best upper bounds are barely below PSPACE. On the other hand,
we do not know of any lower bounds, so in principle they could all be
in P (though this is very doubtful).
The Nash equilibrium problem for 3 players is complete for FIXP;
it is complete in all senses, e.g., its approximation problem is as hard as 
the approximation of any other FIXP problem, the decision problem is
at least as hard as the decision problem for any problem in FIXP, etc. \cite{EY4}.
The price equilibrium problem for algebraic excess
demand functions is another complete problem.

A consequence of the completeness results
is that the class FIXP stays the same under several variations.
For example, using formulas instead
of circuits in the representation of the functions 
does not affect the class (because Nash's function
is given by a formula). 
Also, FIXP stays the same if we use
circuits over $\{+,*, \max \}$ and rational constants (i.e., no division),
because there is another function whose fixed points
are also the Nash equilibria, 
and which can be implemented without division \cite{EY4}.

Of course FIXP contains PPAD, since it contains its complete problems,
for example 2-player Nash. Actually, the piecewise linear fragment
of FIXP corresponds exactly to PPAD. Let {\em Linear-FIXP} be the
class of problems that can be expressed as (reduced to) exact fixed point 
problems for functions given by algebraic circuits
using $\{+,-,\max,\min\}$ (equivalently, \{+,max\}) 
and multiplication with rational constants only; no division
or multiplications of two gates/inputs is allowed.
Then Linear-FIXP is equal to PPAD.

In several problems, we want a particular fixed point of a system $x=F(x)$,
not just any one. In particular, in several of the probems
discussed in Section 3 for example, the function $F$ is a monotone operator
and we want a Least Fixed Point. To place such a problem in FIXP,
one has to restrict the domain in a suitable, but polynomial, way 
so that only the desired fixed point is left in the domain.
For some problems, we know how to  do this (for example, 
extinction probabilities of branching processes), but for others 
(e.g. recursive Markov chains) it is not clear that this can be done in polynomial time.
In any case, the paradigm of a LFP of a monotone operator is
one that appears in many common settings, and which deserves its own separate treatment.

\section{Conclusions}

Many problems, from a broad, diverse range of areas, involve the
computation of an equilibrium or fixed point of some kind.
There is a long line of research 
(both mathematical and algorithmic) in each of these areas, but for many of these
basic problems we still do not have polynomial time algorithms,
nor do we have hard evidence of intractability (such as NP-hardness).
We reviewed a number of such problems here,
and we discussed three complexity classes, PLS, PPAD and FIXP, 
that capture essential aspects of several types of such problems.
The classes PLS and PPAD lie somewhere between P and TFNP (total search problems in NP),
and FIXP (more precisely, the associated discrete problems) lie between P and PSPACE.
These, and the obvious containment PPAD $\subseteq$ FIXP, are
the only relationships we currently know between these classes
and the other standard complexity classes.
It would be very interesting and important to improve on this state of knowledge.
Furthermore, there are several important problems that are in these classes,
but are not (known to be) complete, so it is possible
that one can make progress on them, 
without resolving the relation of the classes themselves.

\newpage
\null

\begin{thebibliography}{Kos97}

\bibitem{ARV}
H. Ackermann, H. Roglin, B. Vocking.
\newblock On the impact of combinatorial structure on congestion games.
\newblock {\em Proc. 47th IEEE FOCS}, 2006.

\bibitem{Al+}
E.~Allender, P.~B\"{u}rgisser, J.~Kjeldgaard-Pedersen, and P.~B. Miltersen.
\newblock On the complexity of numerical analysis.
\newblock {\em Proc. 21st IEEE Comp. Compl. Conf.}, 2006.

\bibitem{AD}
K.J. Arrow, G. Debreu.
Existence of an equilibrium for a competitive economy.
\newblock {\em Econometrica}, 22, pp. 265-290, 1954.


\bibitem{An}
R. M. Anderson.
``Almost" implies ``Near".
{\em Trans. Am. Math. Soc.}, 296, pp. 229-237, 1986.

\bibitem{AH}
R.J. Aumann, S. Hart (eds.).
\newblock {\em Handbook of Game Theory}, vol. 3, North-Holland, 2002.


\bibitem{BCSS}
L.~Blum, F.~Cucker, M.~Shub, and S.~Smale.
\newblock {\em Complexity and Real Computation}.
\newblock Springer-Verlag, 1998.

\bibitem{Ca}
J.~Canny.
\newblock Some algebraic and geometric computations in {PSPACE}.
\newblock {\em Proc. ACM STOC}, pp. 460--467, 1988.

\bibitem{CD1}
X.~Chen and X.~Deng.
\newblock Settling the complexity of two-player {Nash} equilibrium.
\newblock {\em Proc. 47th IEEE FOCS}, pp. 261--272, 2006.

\bibitem{CD2}
X. Chen, X. Deng.
On the complexity of 2d discrete fixed point problem.
{\em Proc. ICALP}, pp. 489-599, 2006.

\bibitem{CD3}
X. Chen, X. Deng.
On algorithms for discrete and approximate Brouwer fixed points.
{\em Proc. ACM STOC}, pp. 323-330, 2005.


\bibitem{CDT}
X.~Chen, X.~Deng, and S.~H. Teng.
\newblock Computing {Nash} equilibria: approximation and smoothed complexity.
\newblock {\em Proc. 47th IEEE FOCS}, pp. 603--612, 2006.

\bibitem{CSVY}
B. Codenotti, A. Saberi, K. Varadarajan, Y. Ye.
\newblock Leontieff economies encode nonzero sum two-player games.
\newblock {\em Proc. SIAM SODA}, pp. 659-667, 2006.

\bibitem{Co}
A.~Condon.
\newblock The complexity of stochastic games.
\newblock {\em Inf. \& Comp.}, 96(2):203--224, 1992.

\bibitem{DFP}
C.~Daskalakis, A.~Fabrikant, and C.~Papadimitriou.
\newblock The game world is flat: The complexity of Nash equilibria
in succinct games.
\newblock {\em Proc. ICALP}, 2006.


\bibitem{DGP}
C.~Daskalakis, P.~Goldberg, and C.~Papadimitriou.
\newblock The complexity of computing a {Nash} equilibrium.
\newblock {\em Proc. ACM STOC}, pp. 71--78, 2006.

\bibitem{dAHK}
L.~de~Alfaro, T. A. Henzinger, O. Kupferman.
Concurrent reachability games.
{\em Proc. IEEE FOCS}, pp. 564-575, 1998.

\bibitem{De}
G. Debreu.
\newblock Excess demand functions.
\newblock {\em J. Math. Econ.} 1, pp. 15-21, 1974.


\bibitem{EM}
A. Ehrenfeucht, J. Mycielski.
Positional strategies for mean payoff games.
{\em Intl. J. Game Theory}, 8, pp. 109-113, 1979.

\bibitem{EB}
P. van Emde Boas.
Machine models and simulations.
In {\em Handbook of Theoretical Computer Science}, vol. A,
J. van Leeuwen ed., MIT Press, pp. 1--66, 1990.


\bibitem{EJ}
E. A. Emerson, C. Jutla.
Tree automata, $\mu$-calculus and determinacy.
{\em Proc. IEEE FOCS}, pp. 368-377, 1991.

\bibitem{EKM}
J.~Esparza, A.~Ku\v{c}era, and R.~Mayr.
\newblock Model checking probabilistic pushdown automata.
\newblock {\em Proc. of 19th IEEE LICS'04}, 2004.


\bibitem{EY1}
K.~Etessami and M.~Yannakakis.
\newblock Recursive Markov chains, stochastic grammars, and monotone systems of
  non-linear equations.
\newblock {\em Proc. STACS}, 2005.
\newblock (Full expanded version available from
\url{http://homepages.inf.ed.ac.uk/kousha}).

\bibitem{EY5}
K.~Etessami and M.~Yannakakis.
\newblock Recursive Markov decision processes and recursive stochastic games.
\newblock {\em Proc. 32nd ICALP}, 2005.

\bibitem{EY2}
K.~Etessami and M.~Yannakakis.
\newblock Efficient qualitative analysis of classes of recursive Markov
  decision processes and simple stochastic games.
\newblock {\em Proc. 23rd STACS}, Springer, 2006.

\bibitem{EY3}
K.~Etessami and M.~Yannakakis.
\newblock Recursive concurrent stochastic games.
{\em Proc. 33rd ICALP}, 2006.

\bibitem{EY4}
K.~Etessami and M.~Yannakakis.
\newblock On the complexity of Nash equilibria and other fixed points.
{\em Proc. IEEE FOCS}, 2007.

\bibitem{FPT}
A. Fabrikant, C.H. Papadimitriou, K. Talwar.
\newblock The complexity of pure Nash equilibria.
\newblock {\em Proc. ACM STOC}, pp. 604-612, 2004.

\bibitem{GGJ}
M.~R. Garey, R.~L. Graham, and D.~S. Johnson.
\newblock Some {NP}-complete geometric problems.
\newblock {\em Proc. 8th ACM STOC}, pp. 10--22, 1976.

\bibitem{Ge}
J.~Geanakoplos.
\newblock {Nash} and {Walras} equilibrium via {Brouwer}.
\newblock {\em Economic Theory}, 21:585--603, 2003.


\bibitem{GZ}
I.~Gilboa and E.~Zemel.
\newblock {Nash} and correlated equilibria: some complexity considerations.
\newblock {\em Games and Economic Behavior}, 1:80--93, 1989.


\bibitem{HJV}
P. Haccou, P. Jagers, V. A Vatutin.
\newblock {\em Branching Processes: Variation, Growth, and Extinction of Populations}.
\newblock Cambridge U. Press, 2005.


\bibitem{Ha}
T.~E. Harris.
\newblock {\em The Theory of Branching Processes}.
\newblock Springer-Verlag, 1963.


\bibitem{HPV}
M. D. Hirsch, C. H. Papadimitriou, S. A. Vavasis.
\newblock Exponential lower bounds for finding Brouwer fixed points.
{\em J. Complexity}, 5, pp. 379-416, 1989.

\bibitem{Ho}
J. J. Hopfield.
\newblock Neural networks and physical systems with emergent 
collective computational abilities.
\newblock {\em Proc. Nat. Acad. Sci.} 79, pp. 2554-2558, 1982.

\bibitem{Jo}
D. S. Johnson.
The NP-completeness column: Finding needles in haystacks.
\newblock {\em ACM Trans. Algorithms} 3, 2007.

\bibitem{JPY}
D. S. Johnson, C. H. Papadimitriou, M. Yannakakis.
How easy is local search?
\newblock {\em J. Comp. Sys. Sci.}, 37, pp. 79-100, 1988.


\bibitem{Ju}
B.~Juba.
\newblock On the hardness of simple stochastic games.
\newblock Master's thesis, CMU, 2005.

\bibitem{Jur}
M. Jurdzinski.
\newblock Deciding the winner in parity games is in UP$\cap$coUP.
\newblock {\em Inf. Proc. Let.} 68, pp. 119-124, 1998.

\bibitem{Ke}
M. Kearns.
\newblock Graphical games.
\newblock In \cite{NRTV}, pp. 159--180, 2007.

\bibitem{Ko}
K.-I. Ko.
Computational complexity of fixpoints and intersection points.
\newblock {\em J. Complexity}, 11, pp. 265-292, 1995.


\bibitem{KS}
A.~N. Kolmogorov and B.~A. Sevastyanov.
\newblock The calculation of final probabilities for branching random
  processes.
\newblock {\em Doklady}, 56:783--786, 1947.
\newblock (Russian).

\bibitem{LH}
C. Lemke, J. Howson.
Equilibrium points of bimatrix games.
{\em J. SIAM}, pp. 413-423, 1964.

\bibitem{LMM}
R.J. Lipton, E. Markakis, A. Mehta.
Playing large games using simple strategies.
{\em Proc. ACM Conf. Elec. Comm.}, 36-41, 2003.

\bibitem{LM}
R.J. Lipton, E. Markakis.
Nash equilibria via polynomial equations.
{\em Proc. LATIN}, 2004.


\bibitem{MS}
C.~Manning and H.~Sch\"{u}tze.
\newblock {\em Foundations of Statistical Natural Language Processing}.
\newblock MIT Press, 1999.

\bibitem{Na}
J.~Nash.
\newblock Non-cooperative games.
\newblock {\em Annals of Mathematics}, 54:289--295, 1951.

\bibitem{NS}
A. Neyman and S. Sorin, eds.
\newblock {\em Stochastic Games and Applications}. 
\newblock Kluwer, 2003. 


\bibitem{NRTV}
N. Nisan, T. Roughgarden, E. Tardos, V. Vazirani.
\newblock {\em Algorithmic Game Theory}.
\newblock Cambridge Univ. Press, 2007.

\bibitem{Pa}
C.~Papadimitriou.
\newblock On the complexity of the parity argument and other inefficient proofs
  of existence.
\newblock {\em J. Comput. Syst. Sci.}, 48(3):498--532, 1994.

\bibitem{Pa2}
C.~Papadimitriou.
\newblock The complexity of finding Nash equilibria.
\newblock In \cite{NRTV}, pp. 29-52, 2007.


\bibitem{Pu}
A. Puri.
\newblock Theory of hybrid systems and discrete event systems.
\newblock PhD Thesis, UC Berkeley, 1995.

\bibitem{Re}
J.~Renegar.
\newblock On the computational complexity and geometry of the first-order
  theory of the reals, parts {I-III}.
\newblock {\em J. Symb. Comp.}, 13(3):255--352, 1992.

\bibitem{RW}
M. Richter, K.-C. Wong.
\newblock Non-computability of competitive equilibrium.
\newblock {\em Economic Theory}, 14, pp. 1-27, 1999.


\bibitem{Ro}
R. W. Rosenthal.
\newblock A class of games possessing pure-strategy Nash equilibria.
\newblock {\em Intl. J. Game Theory} 2, pp. 65-67, 1973.

\bibitem{SS}
R. Savani, B. von Stengel.
\newblock Hard to solve bimatrix games.
\newblock {\em Econometrica} 74, pp. 397-429, 2006.

\bibitem{Sc1}
H.~Scarf.
\newblock The approximation of fixed points of a continuous mapping.
\newblock {\em SIAM J. Appl. Math.}, 15:1328--1343, 1967.

\bibitem{Sc2}
H.~Scarf.
\newblock {\em The Computation of Economic Equilibria}.
\newblock Yale University Press, 1973.

\bibitem{SY}
A. Schaffer, M. Yannakakis.
Simple Local Search Problems that are Hard to Solve.
{\em SIAM J. Comp.}, 20, pp. 56-87, 1991.


\bibitem{Sh}
L.S. Shapley.
\newblock Stochastic games.
\newblock {\em Proc. Nat. Acad. Sci.}, 39:1095--1100, 1953.

\bibitem{Si}
K. Sikorski.
{\em Optimal solution of nonlinear equations}, Oxford Univ. Press, 2001.


\bibitem{St}
B. von Stengel.
\newblock Computing equilibria for two-person games.
\newblock In \cite{AH}, pp. 1723-1759, 2002.

\bibitem{Ti}
P.~Tiwari.
\newblock A problem that is easier to solve on the unit-cost algebraic RAM.
\newblock {\em J. of Complexity}, pp. 393--397, 1992.


\bibitem{Uz}
H. Uzawa.
Walras' existence theorem and Brouwer's fixpoint theorem.
{\em Econ. Stud. Quart.}, 13, pp. 59-62, 1962.


\bibitem{Vo}
B. V\"{o}cking.
\newblock Congestion Games: Optimization in Competition.
\newblock {\em Proc. 2nd ACiD}, pp. 9-20, 2006.

\bibitem{Ya1}
M. Yannakakis.
The analysis of local search problems and their heuristics.
{\em Proc. STACS}, pp. 298-311, 1990.

\bibitem{Ya2}
M. Yannakakis.
Computational complexity of local search.
In {\em Local Search in Combinatorial Optimization}, 
E.H.L. Aarts, J.K. Lenstra eds., John Wiley, 1997.

\bibitem{ZP}
U. Zwick, M. S. Paterson.
The complexity of mean payoff games on graphs.
{\em Theoretical Computer Science}, 158, pp. 343-359, 1996.



\end{thebibliography}
\end{document}